\documentclass{aa501}
\usepackage{graphicx}
\begin{document}
   \title{A box-fitting algorithm in the search for periodic transits}

   \author{G. Kov\'acs \inst{1}, S. Zucker\inst{2} \and T. Mazeh
\inst{2}
}

   \institute{Konkoly Observatory, P.O. Box 67, H-1525,
   Budapest, Hungary\\ \email{kovacs@konkoly.hu}
   \and
   Wise Observatory, Tel Aviv University, Tel Aviv, 69978, Israel \\
   \email{shay@wise.tau.ac.il, mazeh@wise1.tau.ac.il}
   }

   \date{Received 28 February 2002 / Accepted 4 April 2002}

   \titlerunning {A box-fitting algorithm}
   \abstract{ We study the statistical characteristics of a
box-fitting algorithm to analyze stellar photometric time series in
the search for periodic transits by extrasolar planets. The algorithm
searches for signals characterized by a periodic alternation between
two discrete levels, with much less time spent at the lower
level. We present numerical as well as analytical results to predict
the possible detection significance at various signal parameters. It
is shown that the crucial parameter is the effective signal-to-noise 
ratio --- the expected depth of the transit divided by the standard 
deviation of the measured photometric average within the transit. 
When this parameter exceeds the value of 6 we can expect a significant 
detection of the transit. We show that the box-fitting algorithm 
performs better than other methods available in the astronomical 
literature, especially for low signal-to-noise ratios.  
 }

   \maketitle
%

%
%

\section{Introduction}
A considerable fraction of the periodic astronomical time series 
can be modeled rather accurately by finite sums of sinusoidal 
components. In general, these Fourier-sums have a single dominant 
component, and therefore the basic method of Discrete Fourier 
Transformation (DFT) has become commonplace in almost all 
applications (e.g., Deeming 1975). When the signal becomes 
distorted by higher harmonics (e.g., light curves of fundamental 
mode RR~Lyrae and $\delta$~Cephei stars), this simple approach 
fails to perform properly, due to leakage of the signal power to 
many higher harmonics. One way to deal with this problem is to 
use a multifrequency Fourier fit for better approximation of the 
signal shape and thereby to increase the algorithm efficiency, 
a method recently suggested by Defa\"y, Deleuil \& Barge (2001) 
for the search for extrasolar planetary transits. Another generally 
accepted approach is the so-called Phase Dispersion Minimization 
(PDM), which searches for the best period that yields the `smoothest' 
folded time series.

Application of variants of the PDM method in the analyses of variable
star observations goes back to earlier times than that of the DFT.
This is primarily because the PDM algorithm does not require the
computation of trigonometric functions, which put a heavy load on
the computers, especially in those early days. The most frequently
cited implementation of the PDM idea is that of Stellingwerf (1978).
However, earlier versions had appeared already in the '60s and early
'70s, like those of Lafler \& Kinman (1965, hereafter the L-K method),
Jurkevich (1971) and Warner \& Robinson (1972, hereafter the W-R method).
Actually, it can be shown that up to a frequency- (or trial period-)
independent constant, the method of Jurkevich (1971) is equivalent
to that of W-R (see, Kov\'acs 1980). Furthermore, without the
additional feature of overlapping bin structure, the method of
Stellingwerf (1978) is equivalent to that of Jurkevich (1971).

The study of the algorithm presented in this paper has been stimulated
by the increasing interest in searching for periodic transits by
extrasolar planets (e.g., Gilliland et al.  2000; Brown \& Charbonneau
2000; Udalski et al.\ 2002), which follow the discovery of the transit
of HD~209458 (Charbonneau et al.\ 2000; Henry et al.\ 2000) by its
planetary companion (Mazeh et al.\ 2000).  Due to
the short duration of the transit relative to the orbital period
(typically less than 5\%), the signal expected is extremely
non-sinusoidal. Considering the shallowness of the transit (typically
less than 2\% in the case of Jupiter-size planets) and the expected
high noise level of the ground-based small telescopes capable of 
performing large-scale surveys (e.g., Borucki et al. 2001), we 
suggest an algorithm that utilizes the special form of the signal.
The algorithm studied here (see also Gilliland et al. 2000 and Udalski
et al.\ 2002) is based on direct Least Squares (LS) fits of step
functions to the folded signal corresponding to various trial periods.
We present numerical simulations as well as analytical considerations
to estimate the ability of the algorithm to detect a faint signal in a
noisy time series. We show that the algorithm performs significantly
better and more efficiently than the published variants of PDM, DFT or
some LS modification of the latter.

%
%

\section{The box-fitting  algorithm}
We assume a strictly periodic signal, with a period $P_0$, that
takes on only two discrete values, $H$ and $L$. The time
spent in the transit phase $L$ is $qP_0$, where the fractional 
transit length $q$ is assumed to be a small number 
($\approx 0.01-0.05$). For any given set of data points, the 
algorithm aims to find the best model with estimators of five 
parameters --- $P_0$, $q$, $L$, $H$ and $t_0$, the epoch of the 
transit. Actually, if we assume the average of the signal is zero, 
we have $H=-Lq/(1-q)$, and the number of parameters of the model 
is reduced to four.

Let us denote the data set by $\{x_i\}$, $i=1,2, ..., n$. Each $x_i$ 
includes an additive zero-mean Gaussian noise with $\sigma_i$ standard
deviation. The noise is presented by assigning to each data point a
weight $w_i$, defined as $w_i=\sigma_i^{-2}[\sum_{j=1}^n
\sigma_j^{-2}]^{-1}$. It is further assumed that $\{w_ix_i\}$ have
a {\it zero} arithmetic average.

For a given trial period we consider a folded time series, which 
is a permutation of the original time series. This series is 
denoted by $\{\tilde{x}_i\}$ and the corresponding weights by
$\{\tilde{w}_i\}$. We fit a step function to the folded time series
with the following parameters:
\begin{itemize}
\item
$\hat L$ \hskip 0.8mm --- the level in $[i_1,i_2]$
\item
$\hat H$            --- the level in $[1,i_1)$ and $(i_2,n]$
\end{itemize}
The relative time spent at level $\hat L$ is characterized by
$r=\sum_{i=i_1}^{i_2}\tilde{w}_i$, i.e., by the sum of the weights
of the data points at level $\hat L$.

For any given $(i_1,i_2)$, we minimize the expression

%
%
\begin{eqnarray}
\cal D & = & \sum_{i=1}^{i_1-1}\tilde{w}_i(\tilde{x}_i-\hat H)^2
         +   \sum_{i=i_2+1}^{n}\tilde{w}_i(\tilde{x}_i-\hat H)^2 \nonumber \\
       & + & \sum_{i=i_1}^{i_2}\tilde{w}_i(\tilde{x}_i-\hat L)^2 \hskip 2mm .
\end{eqnarray}
Minimization of $\cal D$
yields simple weighted arithmetic averages over the proper index regimes
%
%
\begin{eqnarray}
\hat L ={s\over{r}} \hskip 2mm , \hskip 3mm
\hat H =-{s\over{1-r}} \hskip 3mm ,
\end{eqnarray}
where
%
%
\begin{eqnarray}
s=\sum_{i=i_1}^{i_2} \tilde{w}_i\tilde{x}_i \hskip 2mm .
\end{eqnarray}
With these formulae, the average squared deviation of the fit becomes
%
%
\begin{eqnarray}
\cal D & = & \sum_{i=1}^{n}\tilde{w}_i\tilde{x}_i^2 -
{s^2\over{r(1-r)}} \hskip 2mm .
\end{eqnarray}

Once this expression is evaluated, one has to repeat the computation 
with other $(i_1, i_2)$ values and find the absolute minimum of $\cal D$ 
for any given period. The first term on the right hand side of Eq.~(4) 
does not depend on the trial period, and therefore one can use the 
second term alone to characterize the quality of the fit. We define 
the Box-fitting Least Squares (BLS) frequency spectrum by the amount 
of Signal Residue of the time series at any given trial period:
%
%
\begin{eqnarray}
SR = MAX\biggl\{\biggl[
{{s^2(i_1,i_2)}\over{r(i_1,i_2)[1-r(i_1,i_2)]}}\biggr]^{1\over 2}\biggr\}
\hskip 2mm .
\end{eqnarray}
Here, the maximization goes over the values $i_1 = 1,2, ...,
n^{\star}$, while the $i_2$ values satisfy the inequality
$\Delta i_{min}<i_2-i_1<\Delta i_{max}$, where
$\Delta i_{min/max}$ are determined by the minimum/maximum fractional
transit length suspected to be present in the signal. The maximum lower
index $n^{\star}$ depends on $i_2-i_1$ and covers the range of
$[n-\Delta i_{max},n-\Delta i_{min}]$.

The most obvious meaning of $SR$ follows immediately
from Eqs.~(1) and (4). At the maximum value of $SR$, $\cal D$ of
Eq.~(4) is related to the average variance of the noise. By
using the definition of $\delta \equiv H-L$ for the transit depth,
and the corresponding estimates of $H$ and $L$ of Eq.~(2), we find
that $SR$ at the correct test period yields also an estimate of
$\delta$, i.e., $SR=\hat \delta \sqrt{r(1-r)}$.

In a practical implementation of the above procedure\footnote{ A {\sc
fortran}'77 version of the BLS algorithm is accessible at {\it
http://www.konkoly.hu/staff/kovacs.html}}, we suggest to divide the
folded time series into $m$ bins and evaluate $SR$ by using these
binned values. This approach is very efficient computationally, and
yields an {\it exact} LS solution with a time resolution defined by the
number of bins. Although the lower resolution affects the efficiency
of the signal detection, in all interesting cases a good compromise
can be made between computational constraints and the effectiveness of
signal detection (see next section). One can further minimize the
amount of computation by recognizing that for a given period each
$s(i_1,i_2)$ and $r(i_1,i_2)$ can be obtained by adding the
corresponding bin values to the already evaluated functions of smaller
$i_1$ and $i_2$.

%
%
%
   \begin{figure}[h]
   \centering
   \includegraphics[width=90mm]{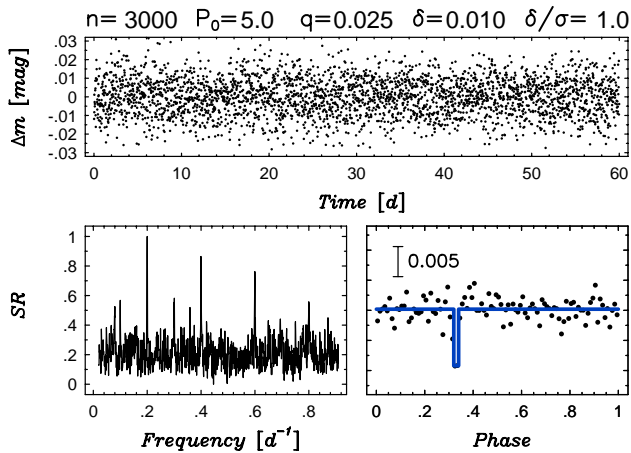}
      \caption{An example of the power of the BLS method. The time
series, the normalized BLS frequency spectrum and the folded time series
are shown in the upper and lower two panels, respectively. The signal
parameters are displayed in the header (see text for details).}
         \label{}
   \end{figure}
As an introductory example of the power of the BLS algorithm, we show
in Fig.~1 the result of the analysis of a test signal with very low
signal-to-noise ratio. Here and in all subsequent simulations we
assume all data points have the same noise level, characterized by
$\sigma$, and therefore the signal-to-noise ratio (SNR) is defined by
$\delta/\sigma$. The figure demonstrates that the BLS spectrum is able
to identify the correct period even at a high noise level.  Several
high peaks appear in the spectrum at integer fractions and multiples
of the true period. This feature is common in all methods utilizing 
higher harmonics of the signal. In this example we used 50
bins, which is a reasonable compromise between computational
efficiency and signal resolution (see next section for details). 
In plotting the folded time series we used 100 bins. Here and in 
all subsequent figures the full frequency band is divided into 
$\approx 1000$ bins and only the maxima in these bins are plotted. 
In this way all important information is retained and showing 
unnecessary details is avoided. The final spectra are normalized 
in the $[0,1]$ interval.

%
%

\section{Properties of the Box-Fitting algorithm}
This section focuses on the applicability of the Box-fitting algorithm
to different time series, and to give signal- and noise-dependent
confidence limits. Due to the statistical complexity, most of the
results are based on extended numerical tests. Nevertheless, whenever
possible, we also present some analytical approximations.

In all subsequent tests the signals have a period of $5^d$ and span a
timebase $T$ of $60^d$. The signal is sampled at times
$t_i=(i-1+\vartheta)T/n$, where $\vartheta$ is a uniformly distributed
random variable in $[0,1]$. This distribution corresponds to the
timings we may obtain during a short but concentrated and
continuous observational campaign from several ground-based
observatories or from space. If the sampling or test periods are
different from the ones used in this paper, the main results presented 
here still remain valid, assuming that the distribution of the data in
the folded time series at any trial period is basically uniform.
Our non-periodic sampling yields aliasing-free spectra in the
frequency band of interest. We do not deal with aliasing effects
originating from nearly periodic sampling, because this problem
affects all period searching algorithms in a similar way.

In all our simulations, the search for the fractional transit
length is performed in the $[0.01,0.10]$ range. The upper limit 
is well above the expected maximum fractional transit length for 
planets (see Defa\"y et al. 2001), but is in the right range for 
detached binaries.

The frequency spectra are computed in the $(0.02,0.91)d^{-1}$ band, a
range that contains frequency components of the true period from the
$10^{th}$ subharmonics up to the $3^{rd}$ harmonics. Usually we use
4000 frequency steps, because, due to the specific signal shape, the
line profiles are very narrow. Therefore, our algorithm requires
a much finer sampling than DFT (see also Stellingwerf 1978).

To characterize the Signal Detection Efficiency we introduce 
(see also Alcock et al. 2000):
%
%
\begin{eqnarray}
SDE & = & {{SR_{\rm peak}-\langle SR \rangle}\over {sd(SR)}}
\hskip 2mm ,
\end{eqnarray}
where $SR_{\rm peak}$ is the $SR$ at the highest peak,
$\langle SR \rangle$ is the average, and $sd(SR)$ is the standard
deviation of $SR$ over the frequency band tested. Because in the  
practical computation of $SDE$ one uses all available spectral 
points, in the presence of periodic signal, the actual value of 
$SDE$ also depends on the time spanned by the data and on the 
lengths and position of the frequency band of the analysis. If 
all other parameters are kept constant, increasing time span or  
frequency band leads to an increase in $SDE$ for signals containing 
periodic component(s). This is because under the condition mentioned, 
the relative contributions to $\langle SR \rangle$ and $sd(SR)$ 
of the peaks associated with the signal become smaller. Of course, 
aliasing leads to a decrease in $SDE$.


\subsection{Dependence on the transit phase}
In the practical implementation of BLS one has to satisfy two
conflicting requirements: (a) a high time resolution ---
demanding a large number of bins; (b) short execution time and
statistical stability --- supporting a small number of
bins. For example, assuming a fractional transit length of 2\%, one
would like to have at least 50 bins, otherwise the transit signal will
be included in a wider bin, degrading the signal. However, 50
bins in such a case are not enough, because in most cases the transit
will be divided into two bins, causing the same effect. For a more
secure coverage we would probably need to double or triple the number
of bins. However, this could be rather time consuming, especially
if the allowed range of fractional transit length is assumed to be
large (say $>0.1$). The sensitive dependence of the execution time
on the number of bins follows from the fact that the number of 
operations in the BLS algorithm is proportional to
$m(\Delta i_{max}-\Delta i_{min})$, and
$\Delta i_{max}-\Delta i_{min}\sim m(r_{max}-r_{min})$.

%
%
   \begin{figure}[h]
   \centering
   \includegraphics[width=85mm]{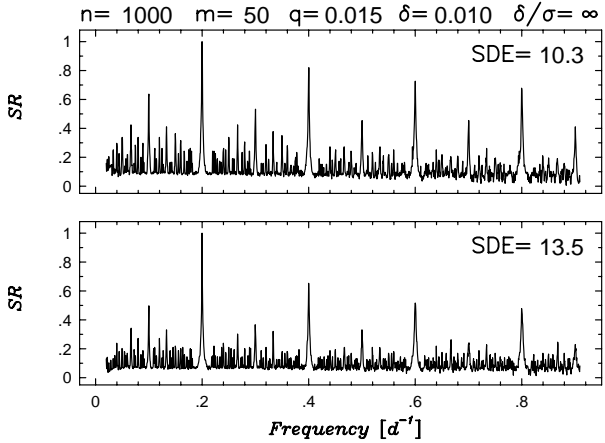}
      \caption{BLS spectra computed for two different transit phases
for the same number of bins and signal parameters, shown in the header.}
         \label{}
   \end{figure}
When using a finite bin size, we expect some dependence of the BLS
spectra on the transit phase, i.e., on the position of the
transit within the folded time series. We illustrate this effect in
the case of a noiseless signal in Fig.~2. The figure shows the strong
increase of the power in the harmonic and subharmonic components, in
addition to similar changes in other substructures of the frequency
spectrum.
%
%
%
   \begin{figure}[h]
   \centering
   \includegraphics[width=85mm]{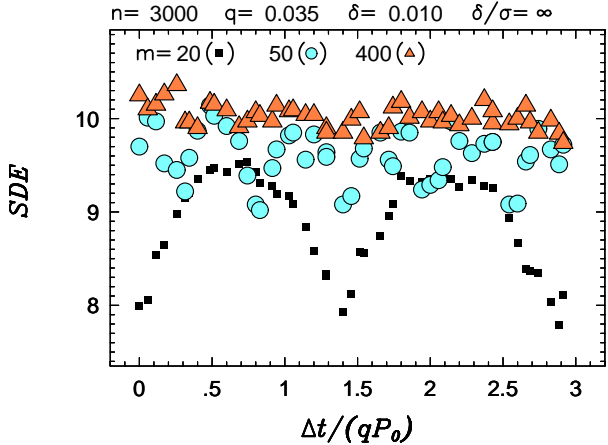}
      \caption{Dependence of the $SDE$ of the BLS spectra on the
transit phase for noiseless signals with parameters shown in the header.}
         \label{}
   \end{figure}
%
%
%
   \begin{figure}[h]
   \centering
   \includegraphics[width=85mm]{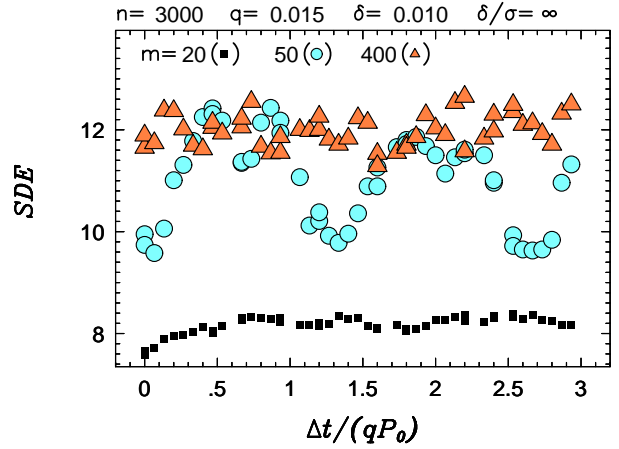}
      \caption{As in Fig.~3, but for a signals of shorter fractional
transit length.}
         \label{}
   \end{figure}

In order to have a better idea of the possible ranges of $SDE$ for 
different numbers of bins, we show in Fig.~3 the $SDE$ as a function 
of the fractional transit phase $\Delta t/(qP_0)$, where $\Delta t$ 
denotes the shift in time of the starting moment of the transit with 
respect to some arbitrary epoch. The transit length is chosen so that 
at $m=50$ the bin size is about half the transit length. The figure
shows that even this number of bins displays relatively large
variations of $SDE$ as a function of the transit phase, let alone
the $m=20$ case. The high number of bins clearly yields high, more
stable $SDE$. However, the necessary increase in the CPU time is
rather high (see later). On the other hand, the low number of bins
may yield a fairly low $SDE$, due to occasional partial coverage of
the transit and wide bin size. The periodicities in $SDE$ come
from the equidistant bin distribution and are given by $(mq)^{-1}$
in units of $qP_0$.

The $SDE$ dependence on the transit phase is a function of the transit
length. To show this function we plot in Fig.~4 the same dependence
for a transit length of $q=0.015$, instead of $0.035$ as in Fig.~3. We
see now that the $SDE$ corresponding to $m=20$ is stable, although
with lower values. This is so because for a small number of bins the full
transit is included in one bin in most cases --- yielding $SR$ values
independent of the transit phase. The strong dependence on the transit
phase occurs when the bin size is comparable to the actual length of
the transit. In both examples, a high enough number of bins yields 
higher, and more stable $SDE$, which, in the presence of noise, means
also a higher probability of signal detection (see Sect. 3.3).

%
%

\subsection{Response to pure noise}

One of the most important features of any search algorithm is its
ability to distinguish between false and true signals. In our case,
this translates to the estimation of the statistical significance of
the highest peak in the BLS spectrum. To study this question we
performed intensive numerical tests to derive the Probability
Distribution Function (PDF) of $SDE$ in case the input data contain
only Gaussian noise.

%
%
%
   \begin{figure}[h]
   \centering
   \includegraphics[width=80mm]{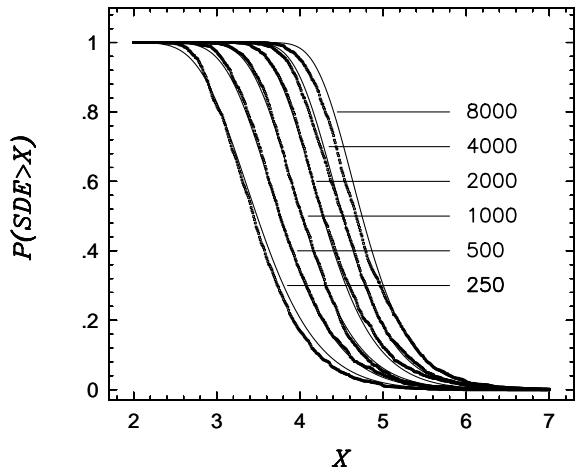}
      \caption{Probability distribution functions of $SDE$ for the BLS
method at various numbers of test frequencies shown at the horizontal
lines. Thick lines are for the empirical (numerical), thin lines are
for the semi-theoretical results described in the text.}
         \label{}
   \end{figure}

Our results show that the PDFs depend only on the
number of trial frequencies $n_f$, and are immune (within the numerical
stability of the simulations) to changing the number of data points
or number of bins used by the BLS algorithm. In order to ensure
reasonable numerical stability, we use in all cases 1500 realizations
to estimate the corresponding PDF. The number of data points is set to
500, 1000, or 2000, while the number of bins is chosen to be 50 or
100. Fig.~5 displays the empirical PDFs obtained for various
numbers of trial frequencies.

One could expect the probability of finding outstanding peaks in a noisy
spectrum to increase with the number of sampled frequencies, if the
samples were statistically independent. We derive a theoretical PDF 
by using this assumption, and show that adjustment of this PDF to 
the empirical one is possible, with parameters different from the 
ones predicted by the assumption of large independent samples.

The computation of $SDE$ consists of two major steps: (a) selection of
the largest $SR(i_1,i_2)$ at each trial period,
(b) selection of $SR_{\rm peak}$ --- the largest $SR$ of all the
values computed for the $n_f$ trial frequencies. In general we have
$mn_f\gg n$, and therefore it is obvious that many of the tested
$SR(i_1,i_2)$ values cannot be considered to be independent
samples. Therefore, in this --- admittedly not exact, but, as we shall
show below, quite practical --- approach, it is assumed that the
distribution of $SDE$ can be approximated by the one obtained for
 $\tilde{n}$ {\it independent} samples of
$SR(i_1,i_2)$, where $\tilde{n}$ is to be determined by numerical
simulations.

Assume, for the sake of simplicity, that all data points have the same
error, $\sigma$. We take $\tilde n$ independent values of the
random variable $SR(i_1,i_2)$, identify the highest value
$SR_{\rm peak}$ and compute $SDE$. For a large sample size we can
assume the
values of $\langle SR \rangle$ and $sd(SR)$ are constant.
Therefore, one can use Eq.~(6) to write the probability that $SDE$
exceeds a specified value $X$:
\begin{eqnarray*}
P(SDE>X) & = & P(SR_{\rm peak} > x) \hskip 2mm ,
\end{eqnarray*}
where $x = X \times sd(SR) + \langle SR \rangle$.
Let $p$ be the probability that a {\it given sample} of $SR$
has a value larger than $x$. Then,
\begin{eqnarray*}
P(SDE>X) =  P(SR_{\rm peak} > x) = 1-(1-p)^{\tilde n}.
\end{eqnarray*}
The value of $p$ can be calculated under the assumption of pure
noise. In such a case $SR$ is actually the absolute value of a zero-mean
Gaussian random variable with a variance of $\sigma^2/n$. This
distribution implies
$sd(SR) = a\sigma/\sqrt{n}$
and
$\langle SR \rangle = b\sigma/\sqrt{n}$
where $a=\sqrt{1-2/\pi}=0.60$ and $b=\sqrt{2/\pi}=0.80$. To facilitate
the calculation of $p$ we can use the normalized random variable
$\chi=SR\sqrt{n}/\sigma$ and write:
%
%
%
\begin{eqnarray}
p &=&  P(SR>x) \nonumber \\
  &=&  P(\chi>x\sqrt{n}/\sigma) \nonumber \\
  &=&  2 (1 - \Phi(x\sqrt{n}/\sigma)) \nonumber \\
  &=&  2 (1 - \Phi(aX+b)) \hskip 2mm ,
\end{eqnarray}
where $\Phi$ is the commulative distribution function of the 
normalized Gaussian variable. If all the $n_f$ samples were 
independent, we could use $n_f$ instead of $\tilde n$. Since 
this is not the case, in order to generalize the above calculation 
we assume that the effective $\tilde n$ is related to $n_f$ by 
some power law: $\tilde n = n_f^c$. The parameters $a$ and $b$ 
depend on the assumption of constant standard deviation and mean 
of $SR$ as well as on its specific distribution. Therefore, in 
order to fit the empirical PDFs we allow the three parameters 
$a$, $b$ and $c$ to vary. The best-fit values we get are: 
$a=0.67$, $b=0.36$, $c=0.83$. These values are significantly 
different from the ones obtained with our simplifying assumptions. 
However, as Fig.~5 shows, the functional form of the above PDF 
with the fitted parameters gives a good approximation of the 
numerical results.

%

\subsection{Signal detection power}
Before turning to the numerical simulations, we derive a simple
estimate of the minimum number of data points necessary obtain a 
high enough signal detection probability. Here we deal with the 
folded time series at the signal period, and ask the question 
whether the differences in $SR^2(i_1,i_2)\equiv
s^2(i_1,i_2)/[r(i_1,i_2)(1-r(i_1,i_2))]$ between the values {\it
including} or {\it excluding} the transit event are significant. We
therefore evaluate the Dip Detection Efficiency:
%
%
\begin{eqnarray}
DDE   & = &
{{E[SR^2(j_1,j_2)] - E[SR^2(k_1,k_2)]}\over
\sqrt{\sigma^2[SR^2(j_1,j_2)]+\sigma^2[SR^2(k_1,k_2)]}}
\hskip 2mm ,
\end{eqnarray}
where the indices $j$, $k$ refer to the `in' and `out' of transit
values, respectively, and $E[...]$, $\sigma^2[...]$ stand for the
expectation value and variance of the corresponding random variable.
If we assume equal noise for all data points and optimal bin size
(i.e., $r=q$), then the averages and standard deviations read:
%
%
\begin{eqnarray}
E[SR^2(i_1,i_2)] & = & S_0 + {\sigma^2\over {n}}
\hskip 2mm , \nonumber \\
\sigma^2[SR^2(i_1,i_2)] & = & 4S_0{\sigma^2\over {n}}
+ 2{\sigma^4\over {n^2}} \hskip 2mm ,
\end{eqnarray}
where $S_0$ denotes the corresponding noise-free value of
$SR^2(i_1,i_2)$, with $i=j$, $S_0=q(1-q)\delta^2$ for the `in transit'
and $i=k$, $S_0=\delta^2 q^3/(1-q)$ for the `out of transit' cases.
By employing the $q\ll 1$ condition and retaining only the leading terms
in $q$, and omitting the quartic terms in $\sigma$, it is easy
to get the final expression for $DDE$
%
%
\begin{eqnarray}
DDE   & = & {1\over 2}{\delta\over \sigma}\sqrt{nq} \hskip 2mm .
\end{eqnarray}
Based on this result, we use
\begin{eqnarray}
\alpha \equiv {\delta\over \sigma}\sqrt{nq}
\end{eqnarray}
to parameterize the {\it effective} SNR, because $\sigma/\sqrt{nq}$ is
the standard deviation of the average of all measurements within the 
transit.

%
%
%
   \begin{figure}[h]
   \centering
   \includegraphics[width=75mm]{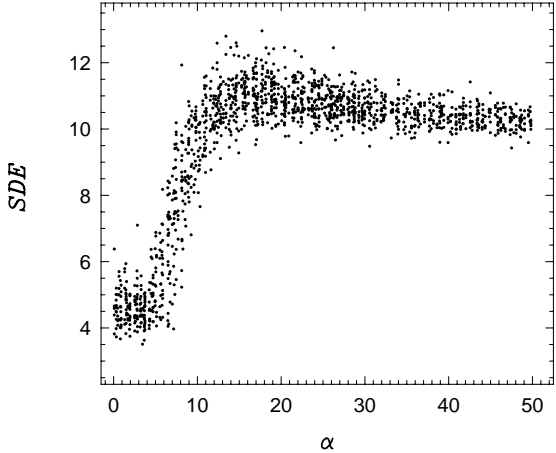}
      \caption{Signal detection efficiency for noisy signals as a
function of $\alpha \equiv {\delta\over \sigma}\sqrt{nq}$. The same
bin number of $100$ and fractional transit length of $0.033$ are used
in all simulations.}
         \label{}
   \end{figure}
We performed $1500$ simulations with $m=100$, $\delta=0.01$, 
$q=0.033$, $n=1000$--$4000$, and various transit phases. The 
dependence of $SDE$ on $\alpha$ is shown in Fig.~6. We see that 
at $\alpha < 5$ the $SDE$ values are noise-dominated at about 
constant level, almost exclusively below 6. The $SDE$ values 
generated by pure noise fall also under $6$ with very high 
probability (see Fig.~5). In the range of $\alpha=6-13$, $SDE$ 
displays a sharp linear increase, until it reaches a mild maximum. 
It turns out that when $\alpha$ gets to the linear increase region, 
the associated value of $SDE$ starts to become significant.

%
%
%
   \begin{figure}[h]
   \centering
   \includegraphics[width=75mm]{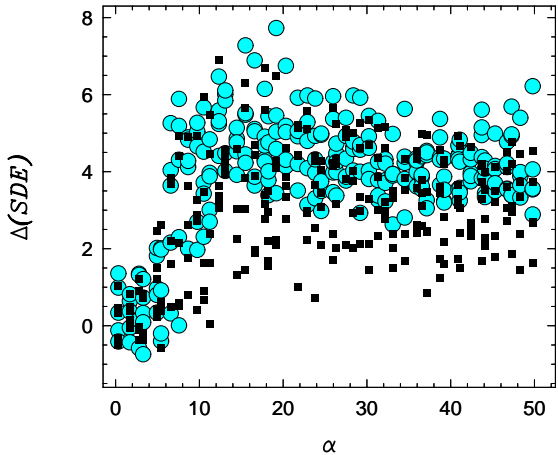}
      \caption{Differential signal detection efficiency for noisy signals
as a function of $\alpha$. Shaded circles and filled squares show results
for $\Delta (SDE)=SDE_{m=200}-SDE_{m=20}$, and $SDE_{m=50}-SDE_{m=20}$,
respectively. The transit length is the same in all cases and is chosen to
be relatively small, i.e., $q=0.015$.}
         \label{}
   \end{figure}

Tests performed with other numbers of bins show that the above pattern
remains the same, with the sole difference that for a larger number of
bins $SDE$ reaches larger maxima, and therefore, the linear
region between $\alpha=6$ and $13$ becomes steeper. It is important to
note that the region of $\alpha$ around 6 is critical in all cases,
because of the separation between the stochastic and deterministic 
signal detections. As we could expect, this regime coincides with 
the $\approx 3\sigma$ limit of $DDE$ (see Eq.~(10)).

In the above simulations we fixed the fractional transit length at a
relatively large value, in order to ensure reasonable resolution with 
$100$ bins. With a shorter transit length, at the same number of bins, 
the $\alpha > 15$ region becomes fuzzier, with a somewhat higher 
average value of $SDE$ (at least for moderately smaller transit 
lengths). However, the global properties remain the same as described 
above.

The response of $SDE$ to changing the number of bins is illustrated 
in Fig.~7. We see that a substantial increase of $SDE$ can be gained 
in many cases by moving from 50 bins to 200 bins. Of course, for 
longer transit lengths, the gain is smaller.


\section{Comparison with other methods}
The purpose of this section is to illustrate that the method
introduced in this paper enables us to discover periodic transit-type
events in noisy time series with a (much) higher probability than the
other standard period searching algorithms available in the current
astronomical literature. In all subsequent examples we use $m=200$
for the BLS method.

%
%
%
   \begin{figure}[h]
   \centering
   \includegraphics[width=75mm]{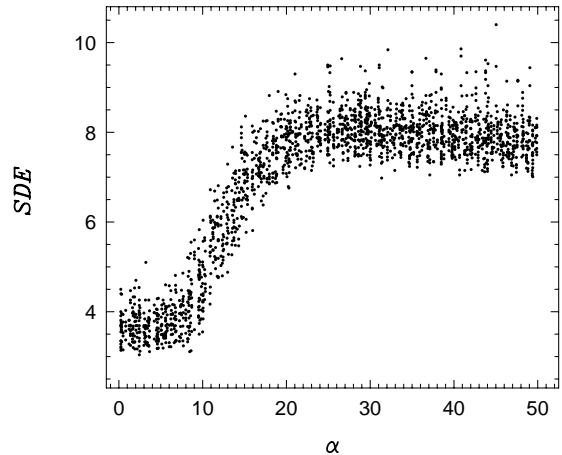}
      \caption{As in Fig.~6, but for the W-R method with $100$ bins.}
         \label{}
   \end{figure}
Perhaps the most competitive method is that of W-R (or similarly, PDM
of Stellingwerf 1978). Therefore, we perform the same test for this
method as the one presented in Fig.~6 for the BLS method. The result
is shown in Fig.~8. For compatibility, we use the same bin number and
signal parameters as for the simulations shown in Fig.~6. In comparison,
we see that the W-R method yields a wider transition region between
the noise- and signal-dominated regimes. Furthermore, the value of
$SDE$ is lower in the signal-dominated region. In order to illustrate
the appearance of these differences in the frequency spectra, we show
a representative example in Fig.~9. We use $100$ bins in the W-R method,
because a larger number leads to an even poorer performance of this
method.
%
%
%
   \begin{figure}[h]
   \centering
   \includegraphics[width=80mm]{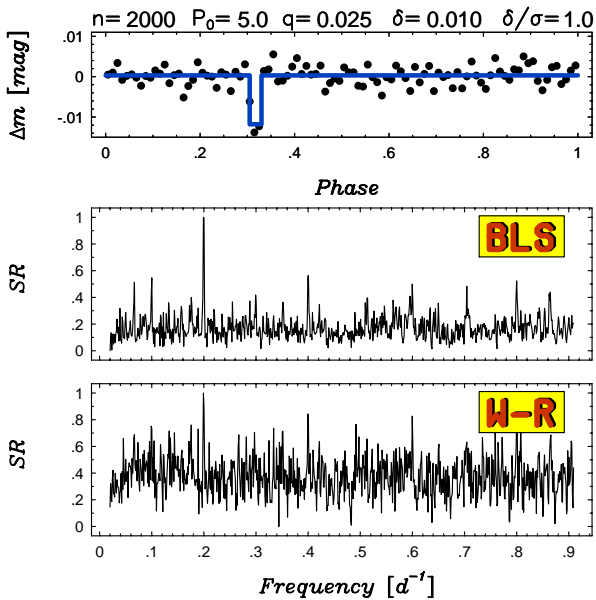}
      \caption{Comparison of the BLS and W-R methods for a realization
of a signal with parameters shown in the header (other parameters are
standard, as given in Sect.~3). The uppermost panel shows the
folded/binned time series (dots) with the period and the fit
(continuous line) obtained by the BLS method.}
         \label{}
   \end{figure}

The next example examines the L-K method. We recall that in
this method the squared differences between the bin-averages are
computed for each folded time series. For large bin numbers and
smoothly varying time series, this method should yield very small
(in the limiting case, zero) L-K statistics at the correct period.
%
%
   \begin{figure}[h]
   \centering
   \includegraphics[width=80mm]{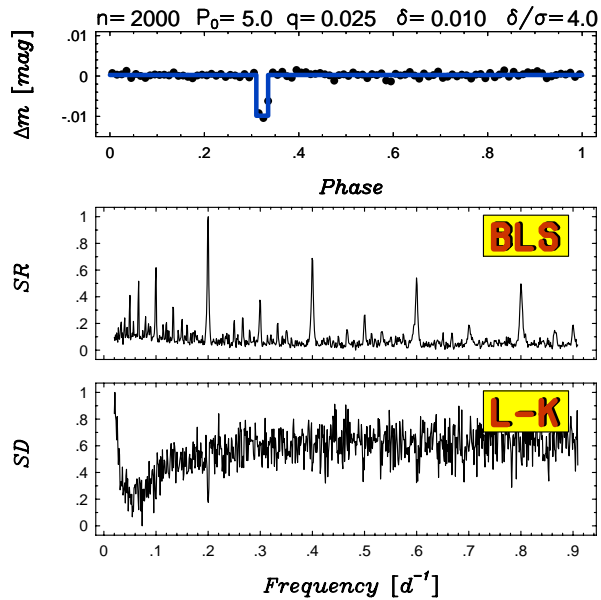}
      \caption{As in Fig.~9, but for the L-K method. On the ordinate
axis for the L-K method, $SD$ stands for `standard deviation', the
corresponding statistics of the L-K method. Minima of $SD$
indicate periodicities in the signal by the the L-K statistics.}
         \label{}
   \end{figure}
Certainly, in the case of periodic signals with discontinuous 
variations, such as the ones studied in this paper, the L-K method 
should give non-zero statistics even in the noiseless case. Indeed, 
even for very high SNR, the L-K method performs extremely poorly 
(see Fig.~10). We use $500$ bins for the L-K method, because tests 
have shown that at lower values the L-K method performs even more 
poorly. By decreasing SNR to 2-3, there remains no significant dip 
close to the test frequency (or to its harmonics) in the L-K spectrum. 
It is important to recall that this noise level still corresponds 
to $\alpha=14$--$21$, well above the secure signal detection limit 
for BLS.

As expected, DFT is also not preferable for analyzing signals
with periodic transits of short duration. In the noiseless case, DFT
yields very slowly decreasing power for the successive harmonics.
Although this is not a good property for correct period identification,
DFT shows a reasonable stability against noise. As shown
in Fig.~11, some remnant power at (or close to) the original harmonics
is still visible in the DFT spectrum, but a reliable identification
is not possible even at this high SNR ($\alpha=11$). At an even higher
SNR, corresponding to $\alpha=14$, the main component becomes the
highest amplitude peak, with a simultaneous increase of the harmonics
and with a still considerable contribution from other parts of the
spectrum, originating from the noise.

%
%
   \begin{figure}[t]
   \centering
   \includegraphics[width=80mm]{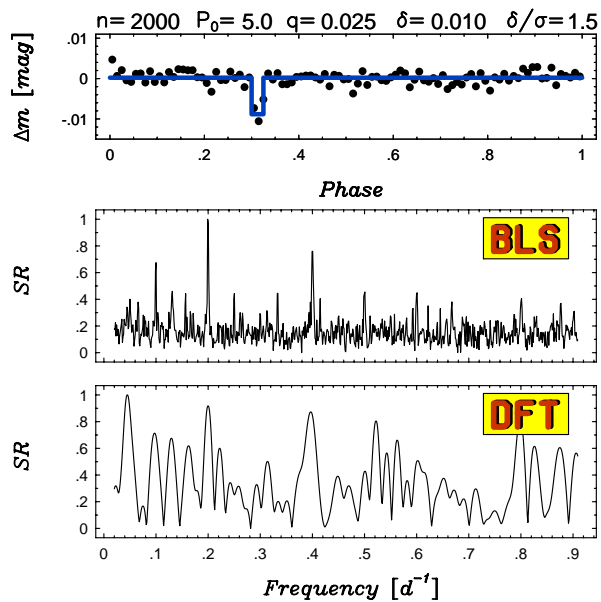}
      \caption{As in Fig.~9, but for the DFT method.}
         \label{}
   \end{figure}

Finally, one can attempt to use multifrequency LS Fourier fit (FLS)
for better approximation of the signal shape and thereby increasing
the $SDE$ in the case of the Fourier method. (We note that for data
distribution leading to orthogonal Fourier base, the method of Defa\"y
et al. (2001) is equivalent to this direct LS Fourier approach.)
Because of the substantial increase in execution time for this method,
we use a smaller number of data points and limit the order of the
Fourier fit to $10$. By plotting the standard deviation of the
residuals, we obtain the result shown in Fig.~12. Although the
performance of FLS could be improved by using more harmonics, FLS did,
in general, a considerably poorer job at moderately high noise levels 
(similarly to the one shown in Fig.~12).

%
%
   \begin{figure}[t]
   \centering
   \includegraphics[width=80mm]{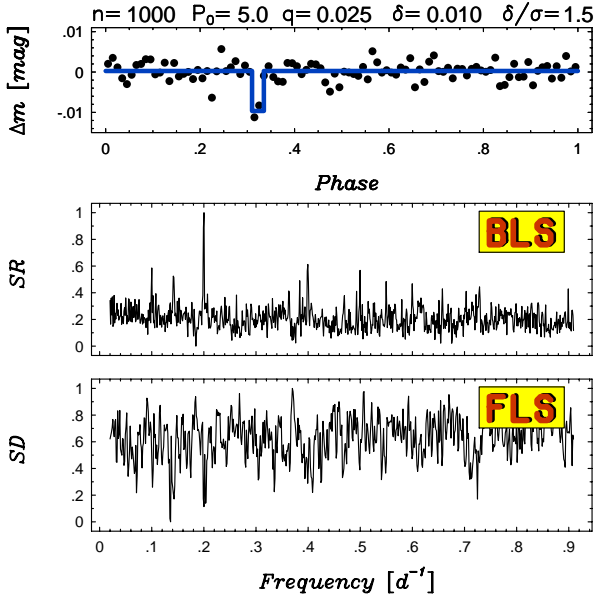}
      \caption{As in Fig.~9, but for the the Fourier-sum-fitting Least
Squares method with 10 components. Minima in the FLS spectrum indicate
periodicities in the signal.}
         \label{}
   \end{figure}
Computational efficiency of the BLS method has been tested in
comparison with DFT on a {\sc sun} workstation. For proper
comparison, the same data sets and the same number of frequency steps
were computed by both methods, although DFT requires about an order of
magnitude fewer steps, because of its wider line
profiles. The range of the search for the best fractional transit
length in the BLS method was fixed at $[0.01,0.10]$, as mentioned at
the beginning of Sect.~3. Table~1 shows that, except
for high bin numbers and low number of data points, BLS surpasses DFT in
execution time per frequency step.  While the CPU time required by DFT is
determined by the number of data points, BLS depends much less on this
parameter. The BLS execution time can be decreased by decreasing the
allowed range of fractional transit length.  For example, with
$n=8000$, $m=500$ the execution time decreases to $13$~s, if 
$[r_{\rm min},r_{\rm max}]=[0.005,0.03]$. Note that by increasing 
the $SDE$ of the Fourier method with the aid of a multifrequency 
least squares approach (or its nearly equivalent version using 
sum of the amplitudes of the harmonics --- see Defa\"y et al. 2001) 
the execution time will increase substantially (proportional to 
the number of harmonics) for the Fourier method. A multifrequency 
Fourier method must use frequency steps as small as for the BLS 
method, while it has to use at least $7$--$10$ harmonics to 
approximate reasonably closely the $SDE$ of the BLS method. 
We therefore suggest that the BLS method can be considered a 
very efficient tool to analyze transit-type signals.

%
%
%
\begin{table}[h]
\caption[ ]{Comparison of the execution times}
\begin{flushleft}
\begin{tabular}{crrc}
\hline
n & m & $t_{\rm BLS}[sec]$ & $t_{\rm DFT}/t_{\rm BLS}$ \\
\hline
1000 &  50  &   1.0 &   6.5  \\
     & 100  &   1.6 &   4.1  \\
     & 200  &   4.0 &   1.6  \\
     & 500  &  19.1 &   0.3  \\
5000 &  50  &   4.3 &   7.6  \\
     & 100  &   5.0 &   6.6  \\
     & 200  &   7.6 &   4.3  \\
     & 500  &  22.6 &   1.5  \\
8000 & 500  &  25.0 &   2.1  \\
\hline
\end{tabular}
\end{flushleft}
{\footnotesize
\underline {Notes:}
$-$~5000 test frequencies are computed \\
\hspace*{25pt}
$-$~a {\sc sun} Ultra 300~MHz machine is used with \\
\hspace*{35pt}
an optimized {\sc fortran} compiler
}
\end{table}
%
%

%
%

\section{Conclusions}
This paper has examined the statistical characteristics of the
Box-fitting Least Squares algorithm to detect periodic transits
in time series of stellar photometric observations. The algorithm
strongly relies on the anticipated box-shape of the periodic light
curve. The advantage of using a predetermined shape of the light
curve manifests itself in the high efficiency of this method relative
to the other search methods, which are generic and can detect any
periodic variation.

The algorithm studied here assumes only two levels of the periodic
light curve. This assumption ignores all other features that are
expected to appear in planetary transits. Thus, we ignore the gradual
ingress and egress phases of the transit, which carry important
information about the parameters of the planetary orbit (e.g, Sackett
1999). The lengths of these phases are short compared to the transit
and thus they are not expected to affect significantly the results of
the search.  Another effect we ignore is the limb-darkening effect,
which has indeed been shown to be small in the case of HD~209458
(e.g., Deeg et al.\ 2001).  The effectiveness of the algorithm relies
on the above simplifying assumption, which is justified as long as we
are interested in a detection tool. After the periodicity is detected
we can try to recover subtle features of the folded light curve, in
order to derive the stellar and the planetary characteristics.

Our main interest is in cases where the signal-to-noise ratio is small, 
and one cannot identify the signal by monitoring a single transit, 
because the stellar drop in intensity is buried in the noise. Contrary 
to the search of transits by the HST in 47 Tuc (Gilliland et al.\ 2000), 
where the noise was small relative to the expected transit dip, we 
concentrate on cases in which the periodic signal can be detected only 
after many measurements are accumulated and the unknown transit is 
monitored many times. To be able to deal with a large number of 
observations, of a thousand or more, we have introduced binning 
into the folded data. We have shown that as long as the bin size is 
small compared to the expected transit length, the efficiency of 
the method is not affected.

One additional factor that determines the computational load of the
algorithm is the range of transit length searched for. The maximum
possible transit length can be estimated if we know the orbital,
stellar and planetary radii. For a given stellar mass, the stellar
radius can be derived by the mass-radius relation, and the orbital
radius can be derived for any period. Recent theories give some
estimates for the planetary radii. Therefore, for a given stellar 
mass we can estimate the maximum duration of the transit, which for
HD~209456 is only a few percent of the period. For most ground-based
and space searches for planetary transits one would have some idea of
the stellar mass of all transit candidates, and therefore we can make
our algorithm computationally more efficient by imposing a variable 
maximum duration on the transit length.

The significance of the detection depends primarily on the effective 
signal-to-noise ratio of the transit. The signal is the stellar
brightness within the transit, relative to the brightness outside the
transit, and the noise is the expected scatter of the measured average
of the stellar brightness inside the transit. The scatter is composed,
obviously, of the observational noise as well as of stochastic
variation of the stellar intensity. It seems that the effective 
signal-to-noise ratio should exceed $6$ in order to get a significant 
detection. This requirement should be taken into account when planning 
future searches for extrasolar planetary transits.

\begin{acknowledgements}
We acknowledge grants {\sc otka} {\sc t--038437}, {\sc t--026031} and
{\sc t--030954} to G.K. and grant 00/40 from the Israeli Science
Foundation.
\end{acknowledgements}

\end{document}